\documentclass[a4paper]{article}
\usepackage{subfigure}
\usepackage{algorithm}
\usepackage{algorithmic}
\usepackage{INTERSPEECH2021}

\title{Improved Conformer-based End-to-End Speech Recognition Using Neural Architecture Search }
\name{Yukun Liu$^{1,2}$, Ta Li$^{1,2}$, Pengyuan Zhang$^{1,2}$, Yonghong Yan$^{1,2,3}$}
\address{
  $^1$Key Laboratory of Speech Acoustics and Content Understanding, Institute of Acoustics, China \\
  $^2$University of Chinese Academy of Sciences, China \\
  $^3$Xinjiang Laboratory of Minority Speech and Language Information Processing, Xinjiang Technical Institute of Physics and Chemistry, Chinese Academy of Sciences, China	}
\email{	liuyukun@hccl.ioa.ac.cn, 
	lita@hccl.ioa.ac.cn,
	zhangpengyuan@hccl.ioa.ac.cn,
	yanyonghong@hccl.ioa.ac.cn}

\begin{document}

\maketitle
\begin{abstract}
Recently neural architecture search(NAS) has been successfully used in image classification, natural language processing, and automatic speech recognition(ASR) tasks for finding the state-of-the-art(SOTA) architectures than those human-designed architectures.
NAS can derive a  SOTA and data-specific architecture over validation data from a pre-defined search space with a search algorithm.
Inspired by the success of NAS in ASR tasks, we propose a NAS-based ASR framework containing one search space and one differentiable search algorithm called Differentiable Architecture Search(DARTS).
Our search space follows the convolution-augmented transformer(Conformer) backbone, which is a more expressive ASR architecture than those used in existing NAS-based ASR frameworks.
To improve the performance of our method, a regulation method called Dynamic Search Schedule(DSS) is employed.
On a widely used Mandarin benchmark AISHELL-1, our best-searched architecture outperforms the baseline Conform model significantly with about 11\% CER relative improvement,
and our method is proved to be pretty efficient by the search cost comparisons.
\let\thefootnote\relax\footnotetext{This work is partially supported by the National Key Research and Development Program of China(No. 2020AAA0108002).}

\end{abstract}

\section{Introduction}
\begin{figure*}[t]
	\centering
	\includegraphics[width=0.85\linewidth]{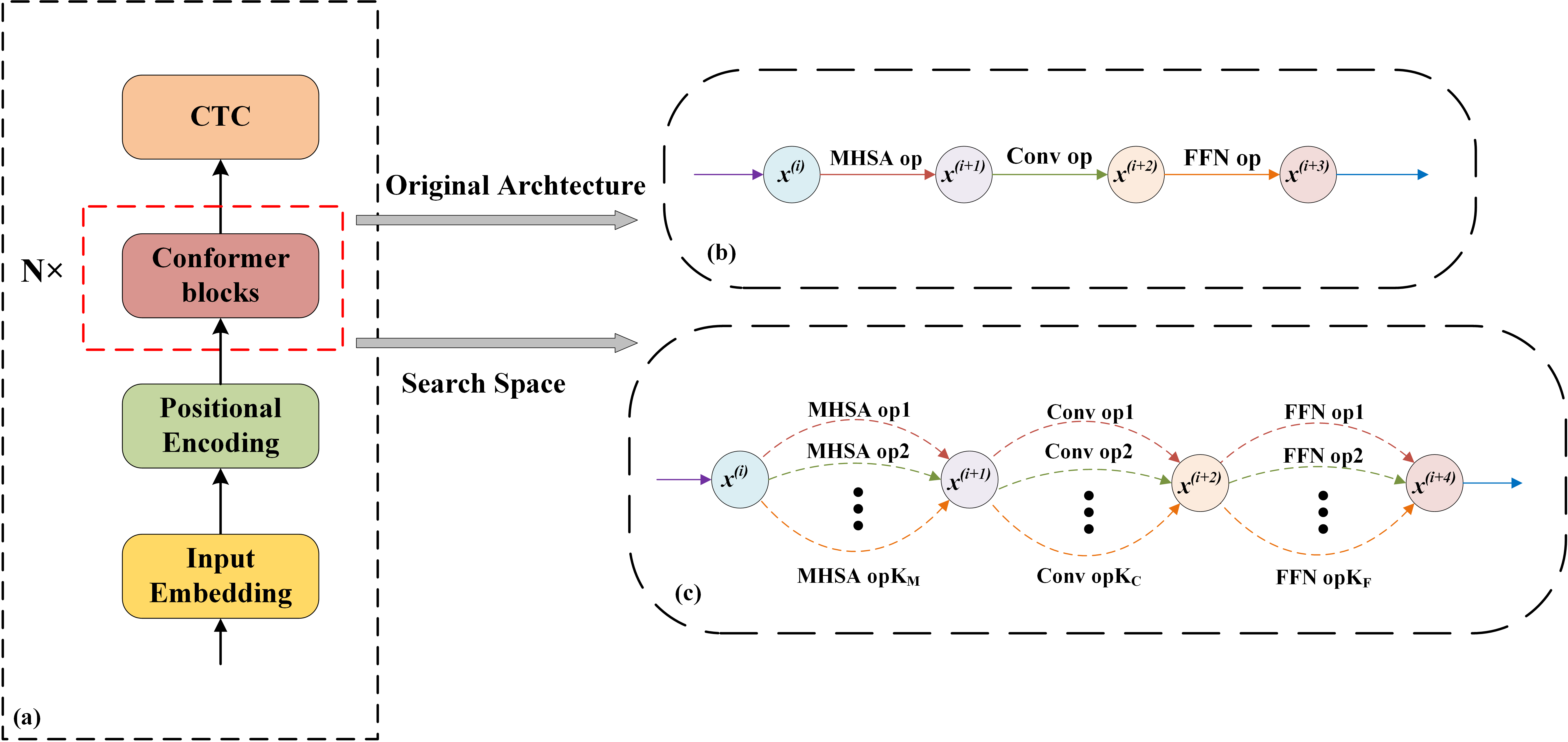}
	\caption{(a) is a overlook of Conformer backbone with N Conformer blocks.
		(b) is the architecture of the original Conformer block, there exist MHSA, Conv and FFN three modules and each module is designed by human.
		(c) is our search space for the Conformer module, where each module can select various operations and  $K_m$, $K_C$, $K_F$ is the candidate operation number of MHSA, Conv, FFN respectively.}
	\label{fig:1}
\end{figure*}

Multi-Headed Self-Attention(MHSA) mechanism has received great success in end-to-end (E2E) models\cite{zhang2020transformer}\cite{ dong2018speech}\cite{zeyer2019comparison} for automatic speech recognition(ASR) tasks.
Different from the recurrent network(RNN) based E2E or hybrid model\cite{graves2012sequence}\cite{rao2017exploring}\cite{chiu2018state}\cite{he2019streaming}\cite{sainath2020streaming}, the MHSA-based model is more easily computed parallelly and has greater power in capturing long-range dependencies, this helps MHSA-based models achieve better performance than the RNN-based.

On the other hand, the convolution(Conv) network also has been successfully applied in ASR tasks\cite{li2019jasper}\cite{han2020contextnet}\cite{abdel2014convolutional}.
Benefitting from that Conv architecture can gather local neighborhood information effectively, some works employ Conv as complementation of MHSA\cite{gulati2020conformer}\cite{huang2020conv}, where Conv and MHSA are designed for capturing local and global information respectively.
In \cite{gulati2020conformer}, a sandwich-like architecture called Conformer is proposed for ASR, which uses Conformer blocks composed of MHSA, Conv, feed-forward network(FFN) three modules and achieves SOTA performance in many benchmarks with the effective joint of MHSA and Conv.
Despite the superior performance of Conformer, all the existing Conformer architectures tend to stack the same block together, however, it remains to be seen whether different combinations of various blocks can further improve the performance.
In one word, it is left to be explored whether we can obtain data-specific Conformer-based architectures beyond existing ones.
This problem is dubbed an architecture challenge.

NAS methods are proposed to designing SOTA neural architecture automatically in a pre-defined search space and have been successfully used for image classification and language modeling tasks\cite{zoph2016neural}\cite{real2019regularized}.
A NAS method contains one search space and one search algorithm.
Existing works have succeeded in constructing the search space base on diverse kinds of networks like Conv\cite{zoph2018learning}, RNN\cite{pham2018efficient}, and MHSA network\cite{so2019evolved}, and a comparable or better architecture than the human-designed can be obtained with the search algorithm over the search space.
Early NAS approaches search architectures with thousands of architecture evaluations, which are computationally expensive, e.g. Reinforcement Learning (RL) based methods \cite{zoph2018learning}\cite{pham2018efficient} and Evolutionary Algorithm (EA) based methods \cite{real2019regularized}\cite{so2019evolved}.
Differentiable Architecture Search(DARTS)\cite{liu2018darts} employed a differentiable gradient-based method that improves the complexity of the neural architecture search effectively. 
With a continuous relaxation of search space, the architecture parameters of DARTS can be optimized directly by gradient descent, this obviously reduces search cost and has discovered SOTA model architectures on many tasks\cite{wu2019fbnet}\cite{chen2019progressive}.

Thus far, there have been several NAS-based works for ASR tasks.
\cite{zheng2020efficient} utilized an improved DARTS algorithm to search the integral architecture over a search space composed of Time-Delay Neural Network(TDNN) operations for ASR task, 
\cite{chen2020darts} directly applied DARTS for searching a partial Conv module architecture used in their ASR models over a Conv-based search space.
Although these DARTS-related works have shown improvement on specific model architecture, they lack efforts of searching MHSA-based architecture, which has been proved great power on ASR tasks.
\cite{kim2020evolved} used EA-based methods to search for MHSA-based ASR model architecture and achieved better performance than models designed with human experiences.
Nevertheless, this method exists two weaknesses, it only searches half blocks of the model, with another half still human-designed.
Even worse, the EA method requires thousands of architecture evaluations, which is extremely computationally expensive.
In \cite{so2019evolved} the same method used for natural language processing(NLP) requires about 270 Google’s TPUs to search architectures within an acceptable time.

As aforementioned, existing human-designed Conformer architectures stack the same block and lack exploration for better combinations.
For existing NAS-based ASR methods, either the backbones of search space\cite{zheng2020efficient}\cite{chen2020darts} are less expressive than Conformer\cite{gulati2020conformer} or the search algorithm is resource comsuming\cite{kim2020evolved}.
To address these problems, we propose a NAS-based ASR framework to search Conformer backbone architecture efficiently, which includes a search space and a search algorithm.
Our search space is also composed of MHSA, Conv, FFN three modules, whereas each module can select various candidate operations.
The widely used DARTS algorithm is employed as our search algorithm, which can search architectures much more efficiently than RL or EA methods, and an additional regulation method called Dynamic Scheduled Search(DSS) is used for further improving performance.
On the AISHELL-1 benchmark, the best architecture found in our experiments outperforms the original baseline model designed by humans with about 11\% CER relative improvement, meanwhile, extensive experiments also prove the efficiency of our method strongly.

%

\section{Related Work}



\subsection{Conformer Architecture}
\label{sec21}
Original Conformer architecture contains an input embedding layer, a positional encoding layer, and a number of Conformer blocks, Figure 1(a) displays an overlook of Conformer architecture.
In the input embedding and positional encoding layer, the input feature will be embedded into the attention dimension and added with the position encoding sequence, which injects positional information into the computation flow.
Then the positional embedding result will be feed into the next Conformer blocks.

There exists three kinds of modules in a conformer block: an MHSA module, a Conv module, and an FFN module, as illustrated in  Figure \ref{fig:1}(b). 
The Conv module begins with a gating mechanism\cite{dauphin2017language} including a pointwise convolution and a gated linear unit, then followed by a 1-D depthwise convolution layer, a batchnorm layer, and a pointwise convolution.
The MHSA module uses the architecture mentioned in \cite{vaswani2017attention}, modified by the relative sinusoidal positional encoding scheme\cite{dai2019transformer} and FFN employs two linear transformations and a nonlinear activation in between.
Residual connection and dropout are applied in all three modules, the same approach is used in many transformer ASR models\cite{dong2018speech}\cite{gulati2020conformer}\cite{vaswani2017attention}.

\subsection{Differentiable Architecture Search}
\label{sec22} 
DARTS is proposed to solve the search efficiency problem caused by RL or EA methods, with a continuous  relaxation.
In DARTS, the search space is represented as a directed acyclic graph(DAG) which consists of nodes and edges, and the DAG is often called the super-network.
As shown in Figure \ref{fig:1}(c), each node $x^{(i)}$ is a latent representation (e.g. a feature map in Conv networks) and each directed edge$(i,j)$ is associated with a candidate operation $o^{(i,j)}$ that can transform $x^{(i)}$ into $x^{(j)}$.

Assuming there exist $K$ kinds different candidate operations $o^{(i,j)}$ between $x^{(i)}$ and $x^{(j)}$, an architecture parameter vector $\alpha^{(i,j)} \in \mathcal{R}^K$ is used to associate with these candidate operations and $\alpha_k^{(i,j)}$ represents the $k$-th operation between $x^{(i)}$ and $x^{(j)}$.
To make the search space continuous, DARTS relaxes the categorical choice of a particular operation to a softmax over all possible operations and use a mixed operation $\bar{o}^{(i,j)}$ to represent the data transformation in the search stage:
\begin{equation}
	\bar{o}^{(i,j)}(x^{(i)}) = \sum_{k<K}\frac{\mathrm{exp}(\alpha_k^{(i,j)})}
	{ \sum_{k^{'}<K}\mathrm{exp}(\alpha_{k^{'}}^{(i,j)})}o(x^{(i)})
	\label{eq1}
\end{equation}
\begin{equation}
	x^{(j)} = \sum_{i<j}\bar{o}^{(i,j)}(x^{(i)})
	\label{eq2}
\end{equation}
In general, DARTS aims to solve a bi-level problem, where the validation dataset is used to optimize the architecture parameters $\alpha$, and the training dataset is used to optimize the operation parameters $\omega$.
Respectively denote $\mathcal{L}_{train}$ and $\mathcal{L}_{val}$ as the training and the validation loss,
this implies the bilevel optimization problem with $\alpha$ as the upper-level variable and $\omega$ as the lower-level variable\cite{colson2007overview}: 
\begin{equation}	
	\min_\alpha
	\mathcal{L}_{val}(\omega^{*}(\alpha),\alpha)
	\label{eq3}	
\end{equation}
\begin{equation} 	
    \mathrm{s.t. }\quad \omega^{*}(\alpha)=\mathrm{argmin}_w\mathcal{L}_{train}(\omega,\alpha)
	\label{eq4}
\end{equation}
In practical search, DARTS approximates $\omega^{*}(\alpha)$ by using a one-step training result for reducing the expense of inner optimization.

\section{Methods}

\subsection{Improved Conformer Using DARTS}
\label{sec31}

Vanilla Conformer is constructed  by stacking identical MHSA, Conv, and FFN modules.
To explore better architectures, we propose a Conformer-based search space, which allows different combinations in each block.
The backbone of our search space follows the sandwich-like architecture of Conformer shown in Figure \ref{fig:1}(a).
Considering in the original Conformer module each node only can be connected to the previous one, each node of our search space follows the same connection limitation, thus Equation \ref{eq2} will be modified as:
%
\begin{equation}
	x^{(i+1)} = \bar{o}^{(i,i+1)}(x^{(i)})
	\label{eq5}
\end{equation}

Figure \ref{fig:1}(c) displays the topology of our search space, where there exist multiple candidate operations for each module.
In Table \ref{tab:1} we list all the candidate operations of three modules, these operations are widely used in Conformer or MHSA-based architectures.
MHSA operations have the same attention dimension 256 but use different attention head numbers.
Conv operations vary in kernel and dilation size, the number in the operation name denotes the kernel size, the default dilation is 1, while \emph{dil\_conv} has the dilation 2.
FFN operations have a distinction in hidden dimension, the number in the FFN operation name means the hidden dimension.
Over the search space, we can derive various architectures containing flexible operation combinations and many existing human-designed conformer architectures can be found in this search space.

DARTS is employed as our search algorithm, the supernet including all candidate operations which will be trained with the bilevel optimization schedule.
After the search process is finished, we preserve the operations with the largest weights in each module, from which we obtain the searched architecture.

%

\begin{table}[h]
	\caption{Candidate operations for  MHSA, Conv, and FFN modules.}
	\label{tab:1}
	\centering
	\begin{tabular}{l p{5cm}}
		\toprule
		\textbf{Module Name}      & \textbf{Candidate Operations}                \\
		\midrule
		MHSA                   & \emph{mhsa\_head4}\quad \emph{mhsa\_head8}\quad \emph{mhsa\_head16}\quad                               														  \\
		\midrule
		Convolution                    & \emph{identity}\quad \emph{conv\_7}\quad \emph{conv\_11}\quad  \emph{conv\_15}\quad    \emph{dil\_conv\_7}\quad \emph{dil\_conv\_11}\quad \emph{dil\_conv\_15}\quad                                   	  \\
		\midrule
		FFN              & \emph{ffn\_1024}\quad \emph{ffn\_512}\quad \emph{ffn\_256}\quad                       	  \\
		\bottomrule
	\end{tabular}
\end{table}

\subsection{Dynamic Search Schedule(DSS)}
\label{sec32}

The original DARTS utilizes a one-step training result as an approximation solution for Equation \ref{eq4}, while in the start-up stage this coarse approximation will lead to a huge gap with the precise solution since the candidate operation parameters $\omega$ are badly learned .
To alleviate this problem, we propose a search schedule called Dynamic Search Schedule(DSS).
The key idea of DSS is using a dynamic-steps approximation as the approximate solution for Equation \ref{eq4} instead of one-step approximation, where the approximation steps ${S_a}$ can be dynamically adjusted with the training steps $S$ of  candidate operation parameters  $\omega$.

Considering the operation parameters $\omega$ is trained with the Noam optimizer\cite{vaswani2017attention}, which adopts a warm-up mechanism and the learning rate $lrate$ of $\omega$ is updated as below:
\begin{equation}
	lrate =  d_{model}^{-0.5}\times min(S\times warmup\_steps^{-1.5}, S^{-0.5})
	\label{eq6}
\end{equation}
%

We refer the warm-up mechanism and employ Equation \ref{eq7} to adjust the approximation steps ${S_a}$ dynamically with two stages.
\begin{equation}
	{S_a} = \mathrm{max}(\beta \times(\frac{S-warmup\_steps}
	{warmup\_steps}), 0)^{-0.5}
	\label{eq7}
\end{equation}

For the first $warmup\_steps$ training steps, $\omega$ is extremely underfitting, so we do not update the architecture parameters $\alpha$ on the badly-learned $\omega$ by setting  ${S_a}$ as positive infinity.
Thereafter $lrate$ is decreased proportionally to the inverse square root of the step number, since $\omega$ is gradually learned well.
For the same reason, we also decrease ${S_a}$ by the inverse square root proportion.
On the other hand,  $\alpha$ relates to the model architecture rather than the model size, we remove the modification $d_{model}^{-0.5}$.
The coefficient $\beta$ is used to normalize ${S_a}$ to a proper value.
Then the calculated approximation steps ${S_a}$ is used for updating $\alpha$, and the entire search procedure with DSS is outlined in Alg. \ref{alg1}.
With manually setting ${S_a}$ as one , the search procedure will involute to the original DARTS which uses the one-step approximation schedule.
\begin{algorithm} 
	\caption{ DARTS with DSS } 
	\label{alg1} 
	\begin{algorithmic} 
\STATE Initialize $S$ and $S_0$ with $0$
\WHILE{\emph{not converged}} 
\STATE Calculate ${S_a}$ with Equation \ref{eq7} 
\IF{$S-S_0$ $\geq {S_a}$} 
\STATE Update  $\alpha$ by descending  $\nabla_{\alpha}\mathcal{L}_{val}(\omega,\alpha)$ and update $S_0$ with $S$
\ENDIF 
\STATE Update  $\omega$ by descending  $\nabla_{\omega}\mathcal{L}_{train}(\omega,\alpha)$
\STATE Replace $S$ with $S+1$
\ENDWHILE
\STATE Derive the final architecture based on the learned $\alpha$ 	
\end{algorithmic} 
\end{algorithm}

\section{Experiments}

\subsection{Data}
\label{sec41}
In this work, we evaluate our search space and search algorithm over a public Mandarin speech corpus AISHELL-1\cite{8384449}.
This corpus contains a 150 hours training set, a 20 hours development set, and a 10 hours test set.
We extract acoustic features using 80-dimensional Mel-filterbanks, which are computed over a 25ms window with a 10ms shift, and 3-dimensional pitch features are added extra.
SpecAugment also is used for data augment\cite{park2019specaugment}.

\subsection{Model Implement}
\label{sec42}
We conducted our experiments with the toolkit ESPnet\cite{2018ESPnet}.
A human-designed model is used as our baseline Conformer with the block number N as four,  and each block is composed of  \emph{mhsa\_head4} , \emph{conv\_15} , \emph{ffn\_1024}, following the finetuned architecture of ESPnet over AISHELL-1.
We use three search methods to derive architecture with search space described in Sect. \ref{sec31}.
Random Search directly samples architectures with a uniform distribution randomly, in our experiments such sampling is run 15 times and the best architecture on the validation is selected as the search result. 
Then for Random and DARTS methods, we both retrain the searched architecture from scratch on training set.  

Connectionist Temporal Classification(CTC)\cite{graves2006connectionist} is adopted as the objective function and the Noam optimizer is used to train our models with warm-up steps = 25000, learning rate = 1.0, label smoothing weight = 0.1 and dropout rate = 0.1.
Additionally, we adopted normalization coefficient $\beta$ as 2.0 for DSS and set the architecture learning rate as 0.0003.
All our experiments are conducted on a single Nvidia Titan RTX.

\label{sec43}
\begin{table}[h]
	\caption{Comparisons between the baseline Conformer architecture and searched architectures by our method. 
	All architectures are evaluated with CER over Dev and Test set.
	}
	\label{tab:2}
	\centering
	\begin{tabular}{p{3.5cm} l l l}
		\toprule
		\  Model      & Model Size & Dev  &Test               \\
		\midrule
		Baseline Conformer                    
		&28.90M
		&7.6
		&8.3
		\\

		Random Search                     
		&26.54M
		&7.6
		&8.4
        \\

		DARTS w/o DSS                    
		&28.85M
		&7.2
		&8.0
		\\

		DARTS with DSS                    
		&28.89M
		&6.7
		&7.5
		\\
		\bottomrule
	\end{tabular}
\end{table} 

\subsection{Result}
\subsubsection{Comparisons with Baseline}
\label{sec431}

Table \ref{tab:2} compares the result(CER) of architectures searched by three methods and the Baseline Conformer.
For a fair comparison, we report the model size of each method.
The Random Search result is worse than the other two DARTS methods meanwhile, which implies that the improvement does owe to our search algorithm.

In the case of the DSS regulation algorithm, both two DARTS methods have achieved better performance than the Baseline Conformer, while DARTS with DSS has the best result among all methods.
The improvement of DARTS with DSS upon DARTS w/o DSS demonstrates the effectiveness of DSS regulation.
DARTS with DSS outperforms the Baseline Conformer with 0.9/0.8 CER in Dev/Test, even DARTS w/o DSS also can gain 0.4/0.3  CER improvement against the Baseline Conformer.




\begin{figure}[h]
	\centering
	\subfigure[ ]{
		\begin{minipage}[t]{0.8\linewidth}
			\centering
			\includegraphics[width=\linewidth]{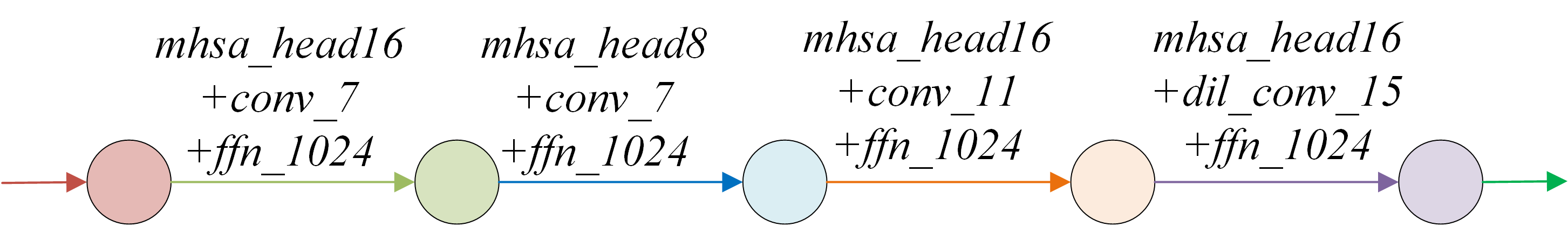}
			\label{fig:2a}
		\end{minipage}%
	}%

	\subfigure[ ]{
	\begin{minipage}[t]{0.8\linewidth}
		\centering
		\includegraphics[width=\linewidth]{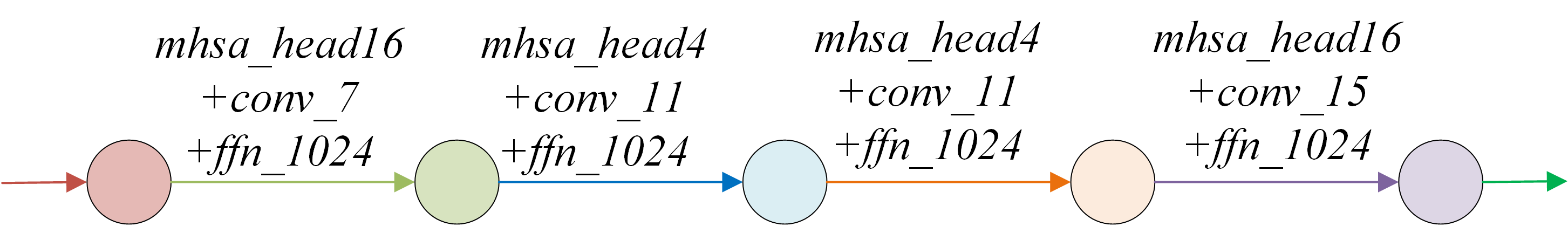}
		\label{fig:2b}
	\end{minipage}%
}%
	\caption{(a) is the best architecture searched by DARTS w/o DSS.
		 (b) is the best architecture searched by DARTS with DSS.}
	\label{fig:2}
\end{figure}

\subsubsection{Searched Architecture}
\label{sec432}

Different from stacking the same blocks of baseline Conformer, our searched architectures can select various operations for each block.
With 4 Conformer blocks, there exist $(3\times7\times3)^{4}=15,752,961$ candidate architectures in our search space.
Figure\ref{fig:2} displays the architectures searched by our methods, corresponding to the performance of Table\ref{tab:2}.

For the MHSA module, \emph{mhsa\_head16}, \emph{mhsa\_head8} are used in Figure \ref{fig:2a} and \emph{mhsa\_head16}, \emph{mhsa\_head4} are used in Figure \ref{fig:2b}.
Operation \emph{ffn\_1024} is selected as the FNN operation for all blocks, which is also used in the Baseline Conformer.
We infer the main reason is that the hidden dimension is crucial for FNN module and a larger dimension always means better performance, thus \emph{ffn\_1024} with the largest hidden dimension is always the best candidate.

Conv operations of two architectures both select larger kernel sizes as the depth increases.
Considering that kernel size is related to the receptive field size of Conv operation, we infer this could be due to the Conv module tends to collect original short-range information in first blocks, then integrate long-range high-level information in last blocks.


\subsubsection{Search Cost Analyze}
\label{sec433}

Table \ref{tab:3} displays the time consumption of our methods.
Search epoch is the epochs of training the supernet and learning architecture, search cost is the total time cost of searching the architecture, and retraining time means the time cost of retraining the obtained architecture from scratch.
Noticing that there exists little difference for the model size of four methods, the retraining times are all extremely close.
Among three search methods, the search cost of Random Search is much longer than another two, since it samples architectures 15 times and each needs to be retrained from scratch to get the validation performance.

\begin{table}[h]
	\caption{Search cost comparison of baseline and our search methods.
		Baseline Conformer has no Search cost since the architecture is human-designed.
	}
	\label{tab:3}
	\centering
	\begin{tabular}{l p{1.3cm} p{1.3cm} p{1.3cm}}
		\toprule
		\ Method  &Search Epoch &Search Cost  &Retraining Time        \\
		\midrule
		
		Baseline Conformer                      
		&-
		&-
		&21.6h
		\\

		Random Search                      
		&-
		&322h
		&21.5h
		\\		
		
		DARTS without DSS                  
		&10
		&17.5h
		&21.6h
		\\
		
		DARTS with DSS                    
		&15
		&23.2h
		&21.6h
		\\
		
		\bottomrule
	\end{tabular}
\end{table} 

From Table \ref{tab:3} we can observe that DARTS methods can reduce search time effectively.
Except for the performance improvement in Sect. \ref{sec431}, DARTS with DSS only needs a little more search cost against DARTS w/o DSS.
Compared with the retraining time, the search cost of DARTS methods is quite acceptable.

\section{Conclusion}
In this paper, we constructed a NAS-based ASR framework.
Our search space was designed based on the Conformer architecture backbone and the widely used DARTS is used as the search algorithm.
Over the mandarin benchmark AISHELL-1, we evaluated our search methods and the searched architectures.
The experiment results showed that the best architecture found by our methods can outperform the baseline architecture significantly.
Furthermore, the comparison of search time cost confirmed the efficiency of our methods.
In future work, we plan to explore the portability of our methods over different E2E architectures.
%

\bibliographystyle{IEEEtran}

\bibliography{mybib}


\end{document}